\documentstyle[aps,draft]{revtex}
\tightenlines
\input{epsf}
\begin{document}
\title{Steady-State Cracks in Viscoelastic Lattice Models}
\author{David A. Kessler\cite{barilan}}
\address{Dept. of Mathematics, Lawrence Berkeley National Laboratory,
1 Cyclotron Road, Berkeley, CA 94720}
\author{Herbert Levine}
\address{Dept. of Physics,
University of California, San Diego La Jolla, CA  92093-0319}
\maketitle
\begin{abstract}
We study the steady-state motion of mode III cracks propagating on a
lattice exhibiting viscoelastic dynamics. The introduction of a Kelvin
viscosity $\eta$ allows for a direct comparison between lattice results
and continuum treatments. Utilizing both numerical and analytical (Wiener-Hopf)
techniques, we explore this comparison as a function of the driving displacement
$\Delta$ and the number of transverse sites $N$. At any $N$, the continuum
theory misses the lattice-trapping phenomenon; this is well-known, but the
introduction of $\eta$ introduces some new twists. More
importantly, for large $N$ even at large $\Delta$,
the standard two-dimensional elastodynamics approach completely misses the
$\eta$-dependent velocity selection, as this selection disappears completely
in the leading order naive continuum limit of the  lattice problem.
\end{abstract}
\pacs{PACS numbers:62.20.Mk, 46.30.Nz}

\section{Introduction}
Recently, there has been renewed interest in the subject of dynamic
fracture~\cite{freund}. This has been sparked in large part by a set of
experiments~\cite{texas,fineberg} that have called into question
some of the predictions of the traditional, continuum mechanics approach
to fracture propagation. Specifically, it has been shown that cracks
exhibit a branching instability long before they reach the predicted
limiting speed of advance; this instability leads to enhanced dissipation
and effectively prevents much additional acceleration. Although there
are some hints of this instability in the continuum approach~\cite{yoffe},
attempts at a systematic analysis~\cite{langer-recent} have been
inconclusive.

In this work, we adopt the philosophy of Marder and Gross~\cite{marder}
and consider lattice models of fracture. These models provide an invaluable
test-bed
for deciding when and if a continuum formulation is appropriate. After all,
if the tip of a brittle crack really occurs at the scale of the lattice, there
is no a priori reason for suspecting that a continuum approach could get the
correct behavior. It is already clear, for example, that lattice models
exhibit a sharp (sometimes discontinuous) jump from static cracks to
propagating ones; this is not reproducible if one neglects lattice scale
effects.
One might hope, though, that at larger velocity there is some effective
continuum description, perhaps utilizing the cohesive zone approach of
Barenblatt~\cite{barenblatt}. From our perspective, it is an open
issue as to whether any such model can accurately predict the behavior of
some specific  microscopic dynamical system exhibiting fracture.

Historically, lattice models of fracture received a major impetus from the
work of Slepyan~\cite{slepyan}, who used the Wiener-Hopf technique to
solve for steady-state propagation. In his work, he considered the
case of infinitesimal dissipation. This fact made it difficult to carry out 
explicit comparisons between
lattice results and continuum predictions thereof, inasmuch as the latter
allows steady-state motion only at the limiting wave speed. One needs
dissipative terms so as to introduce a new macroscopic velocity scale in
order to allow more general steady-state continuum solutions. 
Subsequent analyses by Marder and collaborators~\cite{marder,marder-liu} did 
introduce dissipation in the form of a Stokes term; however, they did not 
explicitly consider the lattice-continuum comparison. In this paper, we
introduce dissipation in a different form by adding a
Kelvin viscosity term to the equation of motion. We will see that the advantage
of this choice is that the continuum model is in fact an accurate 
approximation to the lattice dynamics at large enough $\eta v$. Note too that
from a physics perspective, this type of viscoelasticity appears to have a marked
effect on the crack stability~\cite{sander} and is therefore interesting in its own right.

In this paper, we choose the simplest nonlinear form for the lattice springs,
namely that the spring becomes completely broken (with no residual force)
once it is stretched beyond some threshold. In a future publication, we
will extend our analysis and results to a more general nonlinear force law.
This generalization appears to change very little qualitatively with respect to
the steady-state problem, although it is crucial in allowing for a
direct calculation of the linear stability of the propagating fracture.
As mentioned above, this stability issue is perhaps the one of most
immediate relevance as far as the connection to experiments is concerned.
But, as we shall see, the steady-state problem we solve here offers quite a few
subtle and interesting aspects.

The organization of the paper is as follows.  First, we introduce
the lattice model, and discuss its basic energetic thresholds. We also
briefly discuss the static arrested crack solutions.
In the next section, we generalize the procedure we
employed for finding the static
solutions to solve for steady-state moving cracks for the case of one row
of mass points
($N=1$), employing as a
foundation the Slepyan ansatz for the form of the discrete steady-state
solution.  We analyze the dependence of the velocity on the driving
displacement, studying the effect of the various parameters.  The
most important of these parameters is the viscosity.  In the following section,
we compare these results to a naive continuum limit, finding that
whereas this naive continuum limit successfully reproduces the large-velocity
regime, it fails to account for the nonexistence of solutions below a
driving threshold.  We then extend our method to arbitrary $N$, again
solving for the dependence of the crack velocity on the driving displacement,
as a function of the other parameters.  Subsequently, we
again compare our results to those of our naive continuum limit.  In
particular, we focus on the large $N$ limit, demonstrating how in this
limit the standard continuum calculation is recovered.  We find the surprising
result that whereas the large-scale structure of the displacement field
is almost completely insensitive to the crack velocity for velocities less
than the wave-speed, the small-scale structure is extremely sensitive.  Thus
the whole velocity selection is completely a function of the
lattice-scale dynamics, which no continuum theory can reproduce correctly.
We conclude with directions for the future and some general observations.

\section{Model, Energetics and Statics}
We study in this paper the Slepyan lattice model~\cite{slepyan} for Mode III fracture,
generalized to include Kelvin viscosity.  The model consists of N infinitely
long rows
of mass points (with unit mass) coupled horizontally and vertically 
by damped ``springs''.  The
bottom and top rows are anchored by ``springs'' to lines.  The system
is loaded by extending the top row a distance $\Delta$.  The springs all
have spring constant $1$, except for the bottom row, which has spring 
constant $k$.  All the springs have a viscous damping $\eta$. 
The bottom springs break when their extension exceeds some
threshold $\epsilon$.  We label the (scalar) displacement of the $(i,j)$ mass
from its unstressed equilibrium position as $u_{i,j}$.
The equation of motion for $u_{i,j}$ reads
\begin{equation}
{\ddot u}_{i,j} = (1 + \eta \frac{d}{dt}) (u_{i+1,j} + u_{i-1,j} +
u_{i,j+1} + u_{i,j-1} - 4u_{i,j})
\end{equation}
for $j \ne 1$ with $u_{i,N+1}\equiv\Delta$, and
\begin{equation}
{\ddot u}_{i,1} = (1 + \eta \frac{d}{dt}) (u_{i+1,j} + u_{i-1,j} + u_{i,2}
- 3u_{i,j}) - k \theta(\epsilon - u_{i,1} ) (1+\eta\frac{d}{dt})u_{i,1} \ .
\end{equation}
Of particular interest is the case $k=2$, which is equivalent to the
problem of an up-down symmetric crack, joined at the fracture line by
springs of strength $1$.

There are a number of important strain thresholds which can be understood
from energetics and statics.  The first is the point at which the uniformly
stressed state cracks catastrophically.  For our model in which the
bottom spring has spring constant $k$ and the other vertical springs has
spring constant $1$, for a (horizontally) uniform state, 
the equilibrium displacements are given by
\begin{equation}
u_{i,j} = u_j^U = \frac{(j-1)k + 1}{Nk+1}\Delta \,
\end{equation}
so that the strain of each spring is $k \Delta/(1 + Nk)$, except for the
bottom '$k$' spring which has strain $\Delta/(1 + Nk)$.  The system
will fail catastrophically if the strain of the $k$ spring exceeds 
$\epsilon$ and this gives us our first threshold,
\begin{equation}
\Delta_U = \epsilon (Nk + 1)
\end{equation}

The second strain threshold is the Griffith's criterion, which is
the $\Delta$ at which the uniformly strained state becomes metastable
with respect to the cracked state.
The energy per column of the cracked state,
$u_{i,j}=\Delta$, is just the energy to stretch the $k$ spring to
cracking, $\frac{1}{2}\epsilon^2$, whereas
the energy of the uniformly stressed state is
\begin{equation}
{\cal E}_U = \frac{k\Delta^2}{2(Nk+1)}  \ .
\end{equation}
The cracked state is thus energetically favored when $\Delta$ exceeds
\begin{equation}
\Delta_G = \epsilon \sqrt{Nk + 1}
\end{equation}
Note that this is much smaller than $\Delta_U$ for large $N$.

This system is known to possess stationary solutions which represent
semi-infinite arrested cracks.  For completeness, and to begin to
build the machinery we will need to treat the moving crack, we
briefly outline the solution for this arrested crack. We choose $x=0$ to be
the position of the last uncracked spring.  We solve separately the
problem in the uncracked $(x>0)$ and cracked $(x<0)$ regions and then tie
the answers together.  To solve, we need to know the normal modes of the
vertical springs in the two regions.  We define the general $N \times N$ 
coupling matrix as
\begin{equation} 
{\cal M}_N(m) = \left[
\begin{array}{ccccccc}
 -(m+1)&   1 &     &         &    &    &\\
   1   & -2 &   1  &         &    &    &\\
       &   1 & -2  &   1     &    &    &   \\
       &     &     &  \ddots &    &    &   \\
       &     &     &         &  1 & -2 & 1 \\
       &     &     &         &    &  1 & -2
\end{array}
\right ]
\end{equation}
The coupling matrix on the uncracked side is ${\cal M}_N(k)$ while on
the cracked side it is ${\cal M}_N(0)$.  
Denote the eigenvectors of ${\cal M}_N(0),\ {\cal M}_N(k)$ 
as $\xi_n, \Xi_n$, with eigenvalue $\lambda_n, \Lambda_n$, $i=1,\ldots,N$.
(Here and in the following lower (upper) case symbols refer to quantities 
on the cracked (uncracked) side).

The equation of equilibrium on either side reads
\begin{equation}
0 = u_{i+1,j} - 2u_{i,j} + u_{i-1,j} + {\cal M}_{N;j,j'}(m) u_{i,j'} 
\end{equation}
The general decaying solution on the uncracked side, $i\ge 0$, is
\begin{equation}
u_{i,j}=u^U_j + \sum_n^N A_n (\Gamma_n)^i (\Xi_n)_j
\end{equation}
where $u^U$ is the uniformly strained solution presented above and
\begin{equation}
\Gamma_n=1 - \frac{1}{2}\Lambda_n - \sqrt{-\Lambda_n + 
\frac{1}{4}(\Lambda_n)^2}
\end{equation}
governs  the spatial decay of the $n$th mode, and
satisfies $(\Gamma_i)^2 - (2-\Lambda_i)\Gamma_i + 1 = 0$.
 
The solution on the cracked side, $i\le 0$, is similar:
\begin{equation}
u_{i,j}=\Delta - \sum_n^N a_n (\gamma_n)^i (\xi_n)_j
\end{equation}
where 
\begin{equation}
\gamma_n=1 - \frac{1}{2}\lambda_n + \sqrt{-\lambda_n + 
\frac{1}{4}(\lambda_n)^2}
\end{equation}

This  solution has $2N$ unknowns, $\{A_n, a_n\}$.  The equality
of the two different expressions for $u_{0,j}$ provides $N$ equations.
The equation of motion for $x=0$ provides the other $N$ equations.  Solving
this $2N \times 2N$ inhomogeneous system yields the desired answer.  The
range of validity of this solution is determined by the conditions that
$u_{0,1} < \epsilon < u_{-1,1}$, so that the spring at $i=0$ is the last
unbroken spring.  

Doing this, we find that the possible $\Delta$'s span a range,
$\Delta_A^- < \Delta < \Delta_A^+$ that
encompasses the Griffith's value $\Delta_G$.  
Above $\Delta_A^+$ 
the crack has to run, for it has no other alternative.
Below $\Delta_A^-$, any initial crack would head itself.

\begin{figure}
\centerline{\epsfxsize=3.25in \epsffile{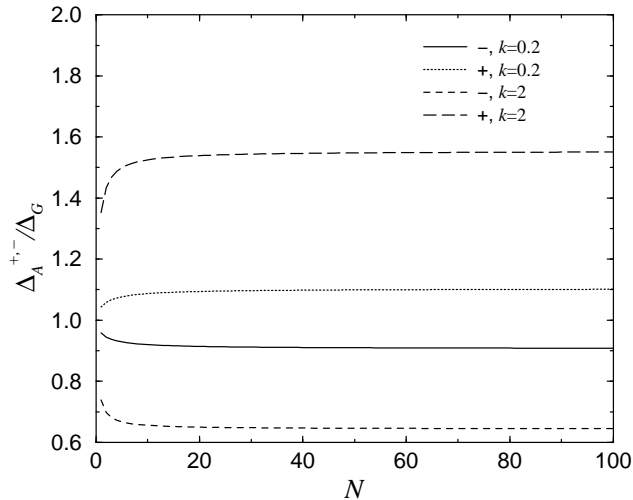}}
\caption{$\Delta_A^\pm/\Delta_G$ vs. N for the case $k=2, 0.2$}
\label{fig1}
\end{figure}

In Fig. \ref{fig1}, we look at $\Delta_A^\pm/\Delta_G$ 
as a function of width for the
natural case,  $k=2$, and the case $k=.2$, where the material has
been weakened along the incipient crack surface. As can be seen,
the effect of the width, $N$, is to widen the window, but the
effect is quite small, once things are normalized to 
$\Delta_G$.  The convergence is numerically consistent with a $O(1/N)$
behavior. We see that $k$, on the other hand, has a dramatic effect, closing
the size of the allowed window significantly.

\section{Moving Cracks, $N=1$}

We now look at moving cracks, starting for simplicity with the case $N=1$.
The key is the Slepyan traveling wave
ansatz
\begin{equation}
u_{i}(t) = u(t-x/v)
\end{equation}
We thus only need to find the function of one variable $u(t)$.  
Plugging this ansatz into
the equation of motion, we get a differential-difference equation which is
non-local in time:
\begin{equation}
\ddot u(t) = (1+\eta \frac{d}{dt})[u(t + 1/v) - 3u(t) + u(t-1/v)] 
- k\theta(\epsilon-u(t)) (1 + \frac{d}{dt})u(t) + \Delta
\end{equation}
We choose $t=0$ to be the moment at which $u$ exceeds $\epsilon$, so 
we can replace the step function above by $\theta(-t)$.
As in the static case, we solve the equation separately in the cracked 
($t>0$) and uncracked ($t<0$) regions. 
It is convenient to discretize time with a small time-step $dt$, so
that $t_i=i\,dt$. Then the 
solution for the uncracked side is
\begin{equation}
u(t_i)=u^U + \sum_l A_l (\Gamma_l)^i 
\end{equation}
where now the $\Gamma$ are given by those roots of
\begin{equation}
\label{poly}
\frac{1}{dt^2}(\Gamma - 2 + \frac{1}{\Gamma}) 
- \left[1+\frac{\eta}{dt}(1-\frac{1}{\Gamma})\right]
\left[\Gamma^{n_b} - 3-k + \Gamma^{-n_b}\right] \ .
\end{equation}
which lie outside the unit circle.  The number $n_b=1/(vdt)$ is constrained
to be an integer, which implies that our resolution in $v$ is limited by
our resolution in $dt$.  There are $2n_b+1$ roots of this polynomial, 
(for $\eta \ne 0$), some
number, $n_u$, of which lie outside the unit circle and thus give rise to
a $u$ which converges as $t \to -\infty$.  
Thus, the
solution for negative $t$ is parameterized by $n_u$ coefficients 
$A_l$,
 $l=\{1,2,\ldots,n_u\}$.  Similarly, we solve in the
cracked region, and the solution is now parameterized by $n_c$ coefficients
$a_l$ corresponding to the roots of
Eq. (\ref{poly}) (with $k$ set to zero) 
which lie inside the unit  circle.  It can also be shown that for sufficiently
small $dt$, $n_u + n_c=2n_b + 1$.  Thus the entire solution is parameterized
by $2 n_b + 1$ parameters. As in the static case, the two solutions
overlap, this time for $2 n_b$ values of $t_i$, $i=-n_b,-n_b+1,\ldots,n_b-1$,
so the last uncracked equation at $i=-1$ fixes $u$ out to $i=n_b-1$ and
the last cracked equation at $i=1$ likewise fixes $u$ down to $i=-n_b$.
The identity of the two expressions for $u$ in the overlap region give us
$2n_b$ equations. The last  equation we need comes from the  equation
of motion at the crack point $i=0$.  Solving this inhomogeneous system gives
us our desired solution.  Reading off $u_1(0)=\epsilon$ gives us the relation
between $v$ and $\Delta/\epsilon$ we need.

\global\firstfigfalse

Again, as in the static problem, there is a consistency constraint on the
solution, namely that $u_1(t)$ not reach $\epsilon$ before $t=0$.  In the
$\eta=0$ problem studied by Marder, this happens for too small $v$.  This
holds true, as we shall see, for sufficiently small $\eta$.

\begin{figure}
\centerline{\epsfxsize=3.25in \epsffile{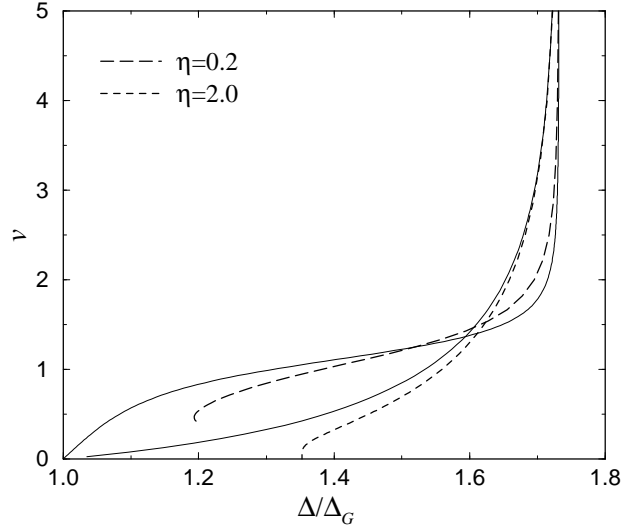}}
\caption{$v$ versus $\Delta/\Delta_G$ for $\eta=0.2,\ 2$ for $N=1$,
$k=2$. The calculation was done with $dt=.05$.  
The solid lines are the naive continuum results for the same
parameters.}
\label{fig2}
\end{figure}

In Fig. \ref{fig2}, we present data on $v$ versus $\Delta/\Delta_G$ for 
two values of $\eta=0.2,\ 2$ with $k=2$.  Two striking features
present themselves.  First is the divergence of the velocity, which
occurs for both values of $\eta$ at $\Delta_U$ which we calculated 
in Sec. 2 as the displacement
for which the entire system wants to break apart.  Note that in this
model, there is nothing wrong with $v>1$, the wave speed in our units. The second important
feature of these curves is the different behaviors exhibited at the
left edge of the graph.  The low $\eta$ graph exhibits
the typical behavior of a subcritical bifurcation which ends at a
square-root type cusp.  The continuation past the cusp to even lower 
velocities is a numerical
artifact of finite $dt$ and vanishes in the $dt \to 0$ limit.  This cusp
feature persists in the $\eta \to 0$ limit, and for smaller velocities
the solutions are inconsistent with the condition mentioned above that
$u(i)<\epsilon$ for $t<0$ and so are unphysical.  For larger $\eta$,
where the system is sufficiently overdamped, the solutions 
persist to zero velocity.
However, the dependence on $\Delta/\Delta_G$ is very singular, and the
graph approaches zero velocity at $\Delta_A^+$ exponentially in $1/v$.
Before proceeding to a full survey of parameter space, 
it is useful to develop a naive continuum
limit for our system, which is analytically more tractable and serves
as a useful benchmark for our discussions.  We do this in the next section.

\section{Naive Continuum Limit}

We now develop the naive continuum limit of
the equation of motion for $N=1$ steady-state moving cracks.  We obtain this limit
by simply replacing the finite difference  by a derivative, yielding:

\begin{equation}
\label{n_1eqofmot}
\ddot u(t) = (1+\eta \frac{d}{dt})\left(\frac{1}{v^2}\frac{d^2u}{dt^2} - u\right)
- k\theta(-t) (1 + \eta\frac{d}{dt}) u + \Delta
\end{equation}

The solution for $t<0$ is
\begin{equation}
u(t) = \frac{\Delta}{1+k} + A_1e^{Q_1 v t}
\end{equation}
where $Q_1$ is the unique root with positive real part of the
polynomial $P(-(1+k);Q_1)$ where in general $P(\lambda;Q)$ is defined by
\begin{equation}
\label{Qpoly}
P(\lambda;Q)\equiv\eta v Q^3 +  (1 - v^2) Q^2 + (1 + \eta v Q)\lambda  \;.
\end{equation}
In general, for $\lambda<0$, $P$ has one positive root, with the other
roots having negative real parts.
Similarly, the solution for $t>0$ is
\begin{equation}
u(t) = \Delta -  a_2 e^{-q_2 v t} - a_3 e^{-q_3 v t}
\end{equation}
where $-q_2$, $-q_3$ are the two roots with negative real part of $P(-1;q)$
One can solve for $A_1$, $a_2$, and $a_3$ by using the continuity of
$u$ and its first two derivatives at $t=0$.

One can learn a few things from these equations.  As pointed out by
Langer in his study of a related model~\cite{langer}, vanishing $\eta$ in the
continuum is a singular limit, since it controls the highest derivative.
Secondly, the special role of the wave speed, $v=1$, is clear.  Most 
interesting,
however, is the question of when we expect this continuum limit to be valid.
The condition is that at least some of the exponential decay rates are small,
such that the solution will look smooth on the lattice scale. For small $v$, $Q_1 \sim \sqrt{k+1}$,
$q_2 \sim 1$, and $q_3 \sim 1/(\eta v)$, so none of the $q$'s are small
and the continuum limit is not reliable.  For large $v$, things are
different.  Here, $Q_1 \sim v/\eta$, $q_{2,3} \sim 
(\eta \pm \sqrt{\eta^2 - 4})/(2v)$.  The large value of $Q_1$ corresponds
to the existence of a boundary layer which allows for the matching
of the highest derivative term.  If this boundary layer does not
affect the lower order matches (as is fairly typical), then we would
expect that the continuum result will agree with the lattice answer.

It is straightforward to work out the small velocity limit of
the continuum theory.  Using the limiting values of the
$q$'s above, and solving the linear system, one gets that $\epsilon \sim 
\Delta /\sqrt{k+1} + O(v)$, so that $\Delta \sim \Delta_G + O(v)$.
Thus, the continuum solution starts at the Griffith's point $\Delta_G$ 
with zero velocity and the velocity grows linearly as $\Delta$ is increased.
Continuing the calculation to first order in $v$, we find 
$v = (\Delta/\Delta_G-1)/[(\sqrt{k+1}-1)\eta]$.
Thus, the velocity is inversely proportional to $\eta$. This is as
expected, since for $v\ll1$, the velocity only enters in the
combination $\eta v$.

The large velocity limit is also analyzable. Again using the $q$'s found
above, we obtain to leading order that $v = \left[(1-\Delta_U)/(k\eta^2)
\right]^{-1/4}$ so that $v$ diverges at $\Delta_U$.  Note also that
in this regime $v$ scales as $\sqrt{\eta}$, so that as viscosity increases
so does the velocity.  This is because the only reason that propagation 
at $v>1$ is possible is because of the viscosity, so the larger the
viscosity the more efficient the propagation.

As $\eta$ goes to zero, the continuum limit must break down.  The velocity
increases very rapidly (at a rate proportional to $1/\eta$) to near $1$, 
and stays there till $\Delta$ is
near $\Delta_U$, whence it rapidly diverges.  The velocity crosses unity
at $\Delta=(1+k)^{2/3}$ with slope of order $\eta^{2/3}$.  Thus, with
vanishingly small $\eta$, steady-state propagation, at least in our
naive continuum limit, is only possible at the wave speed $v=1$.

We are now in a position to compare our lattice calculation to the
continuum results.  For example, Fig. \ref{fig2} above also displays
the continuum curves.  We see that in general the continuum 
calculation is very good
only at the largest velocities.  The agreement gets progressively worse
for smaller velocities and breaks down completely for $\Delta$ less
than the arrest value $\Delta_U^+$.  This must be since the continuum
calculation has the velocity going smoothly to zero at $\Delta_G$, which
it never does due to the existence of arrested solutions. Also, the
agreement is better at larger $\eta$; above some critical $\eta$, (about
$\eta\approx 0.5$ for $k=2$), the
continuum curve serves as an upper bound on the lattice curve and thus
is a fair approximation until we start getting close to the arrest value.
All of this is to be contrasted to what would have been obtained 
had we set $\eta=0$ 
and introduced instead a Stokes velocity term proportional to $\dot{u}$.
Then, the continuum theory has a limiting velocity of $v=1$ and since this
feature is not shared by the lattice dynamics, the continuum approximation
would be nowhere accurate.

\begin{figure}
\centerline{\epsfxsize=3.25in \epsffile{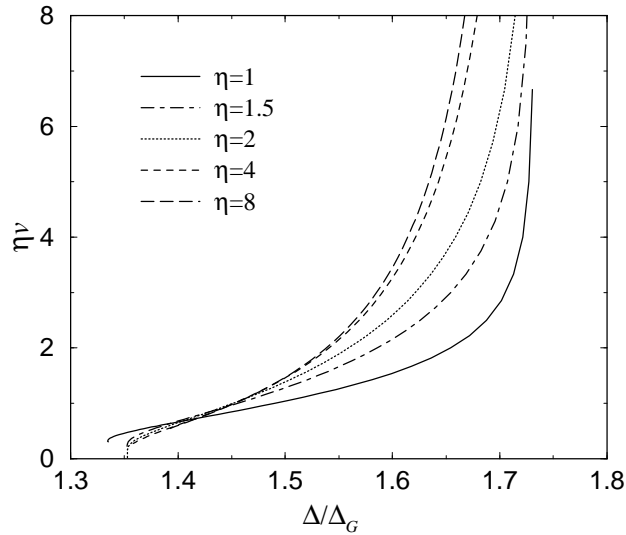}}
\caption{$\eta v$ versus $\Delta/\Delta_G$ for $\eta=1$, 1.5, 2, 4, 8 for $N=1$,
$k=2$. The calculation was done with $dt=.05$.}
\label{fig2a}
\end{figure}

We saw from the continuum calculation that for velocities $v \ll 1$, the
relevant parameter for the continuum calculation is $\eta v$.  We can test
this for the lattice model by plotting $\eta v$ versus $\Delta$.  This is
presented in Fig. \ref{fig2a}.  We see that asymptotically for large
$\eta$ this scaling sets in, but for finite $\eta$, the existence of the
lattice length scale ruins the simple scaling.

\begin{figure}
\centerline{\epsfxsize=3.25in \epsffile{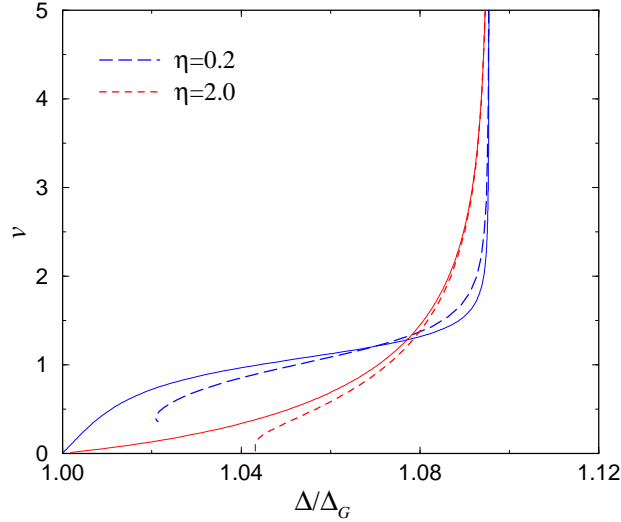}}
\caption{$v$ versus $\Delta/\Delta_G$ for $\eta=0.2,\ 2$ for $N=1$,
$k=0.2$. The calculation was done with $dt=.05$.  
The solid lines are the naive continuum results for the same
parameters.}
\label{fig3}
\end{figure}

It is interesting to see what happens for smaller $k$.  We saw that
the window of arrested solutions is significantly smaller for smaller
$k$.  In Fig. \ref{fig3}, we present the analog of Fig. \ref{fig2},
but this time with $k=0.2$.  We see that the results are as far from
the continuum limit as they were with the larger $k$.  While the
window of arrested solutions is smaller, so is the value of $\Delta_U/
\Delta_G$, so the entire picture is just shrunk to a smaller range of
$\Delta$.

\subsection{Wiener-Hopf Solution}

Before we turn to general $N$, we will present the equivalent
Wiener-Hopf solution of the $N=1$ problem.  This method is more
involved for the case $N=1$, but it is a model for the Wiener-Hopf 
solution we will present for general $N$, which we allow us to draw
analytical conclusions in the large $N$ limit.  The basic method follows
that of Marder and Gross~\cite{marder}, but the presence of viscosity
adds some new twists which are worthy of comment.

To begin, we define the Fourier-transform ${\tilde u}^\pm$ 
of the right- and left-hand pieces of the $u$ field as follows:
\begin{equation}
{\tilde u}^\pm(K) = \int_{-\infty}^{\infty}\,vdt \ \theta(\pm t)
e^{iKvt} u(t)
\end{equation}
It should be noted that ${\tilde u}^\pm$ are analytic in the
upper and lower half-planes respectively.
The Fourier transform of the $u$ field, $\tilde u$, is just the sum of the
two parts: $\tilde u(K)={\tilde u}^+ + {\tilde u}^-$. In terms of these
fields, the equation of motion reads
\begin{equation}
0=\left(i\eta v K^3 - (1 - v^2)K^2\right){\tilde u}
- (1 - i\eta vK){\tilde u} - k(1-i\eta vK){\tilde u}^- +
\Delta\delta(K) - k\eta v u(0)
\end{equation}
The last term is noteworthy, and arises because the time derivative
does not act on the $\theta$-function in the last term in Eq. \ref{n_1eqofmot}.
Expressing $\tilde u$ in terms of its component pieces, we recognize
that the coefficients of ${\tilde u}^\pm$ are nothing put $P(-1;-iK)$,
$P(-(1+k);-iK)$ respectively, that we encountered in our solution above,
so that the the equation of motion reads
\begin{equation}
0=P(-(1+k);-iK){\tilde u}^- + P(-1;-iK){\tilde u}^+
 + \Delta\delta(K) - k\eta v u(0)
\end{equation}
Next, we factor the $P$'s in terms of their roots
\begin{eqnarray}
P(-(1+k);-iK) &=& i\eta v(K - iQ_1)(K+iQ_2)(K+iQ_3) \nonumber \\
P(-1;-iK) &=& i\eta v(K - iq_1)(K+iq_2)(K+iq_3) 
\end{eqnarray}
Dividing through by $i\eta v(K-iq_1)(K+iQ_2)(K+iQ_3)$ we obtain
\begin{equation}
0=\frac{K-iQ_1}{K-iq_1}{\tilde u}^- + \frac{(K+iq_2)(K+iq_3)}{(K+iQ_2)(K+iQ_3)}
{\tilde u}^+ - \frac{\Delta}{\eta vq_1Q_2Q_3}\delta(K) 
+ \frac{ik}{(K-iq_1)(K+iQ_2)(K+iQ_3)}u(0)
\end{equation}
where we have used the $\delta(K)$ to simplify its prefactor.  We
use the fact that $q_1q_2q_3=-P(-1;0)/(\eta v)= 1/\eta v$ to
rewrite our equation as
\begin{equation}
0=\frac{K-iQ_1}{K-iq_1}{\tilde u}^- + \frac{(K+iq_2)(K+iq_3)}{(K+iQ_2)(K+iQ_3)}
{\tilde u}^+ - \frac{\Delta q_2q_3}{Q_2Q_3}\delta(K) 
+ \frac{ik}{(K-iq_1)(K+iQ_2)(K+iQ_3)}u(0)
\end{equation}
To proceed, we have to decompose the two inhomogeneous terms into pieces
analytic in either the upper- or lower- half-planes. The result is
\begin{eqnarray}
0&=&\frac{K-iQ_1}{K-iq_1}{\tilde u}^- + \frac{(K+iq_2)(K+iq_3)}{(K+iQ_2)(K+iQ_3)}
{\tilde u}^+ - \frac{\Delta q_2q_3}{Q_2Q_3}\left(\frac{i}{K+i0^+}
- \frac{i}{K-i0^+}\right) \nonumber\\
&\ &\ + \frac{ik}{(q_1+Q_2)(q_1+Q_3)}\left(-\frac{1}{(K-iq_1)} +
\frac{K+i(q_1+Q_2+Q_3)}{(K+iQ_2)(K+iQ_3)}\right)u(0)
\end{eqnarray}
The key to the Wiener-Hopf method is the realization that the sum of the 
terms analytic in either half-plane have to vanish, allowing us to solve
for ${\tilde u}^\pm$.  We find
\begin{eqnarray}
{\tilde u}^+ &=& \frac{i\Delta q_2q_3 (K+iQ_2)(K+iQ_3)}{Q_2Q_3(K+i0^+)(K+iq_2)(K+iq_3)}
 - \frac{ik(K + i(q_1 + Q_2 + Q_3))}{(q_1+Q_2)(q_1+Q_3)(K+iq_2)(K+iq_3)}u(0)
\nonumber \\
{\tilde u}^- &=& \frac{-i\Delta Q_1 (K-iq_1)}{(1+k)q_1(K-i0^+)(K-iQ_1)}
 + \frac{ik}{(q_1+Q_2)(q_1+Q_3)(K-iQ_1)} u(0)
\end{eqnarray}
Notice that the poles in ${\tilde u}^\pm$ give rise to exactly the
same exponential terms in $u$ that we found previously.  It can be
explicitly verified that the two forms of the solution are equivalent.
For our purposes, it is sufficient to examine what happens in the
small $v$ limit.  Then, as we have already noted $q_3, Q_3 \approx 1/(\eta v)$.
The first, $\Delta$, term on the right-hand sides approaches a finite limit, 
with $O(v^2)$ corrections, whereas
the second $u(0)$ term vanishes linearly in $\eta v$.  More explicitly,
we find
\begin{equation}
{\tilde u}^+ \approx \frac{i\Delta q_2q_3 (K+iQ_2)}{Q_2Q_3(K+i0^+)(K+iq_2)}
 - \frac{ik \eta v}{(q_1+Q_2)(K+iq_2)} u(0)
\end{equation}
Using $q_1=q_2=1$, $Q_1=Q_2=\sqrt{1+k}$, and $q_3=Q_3=1/\eta v$, 
and using the inverse Fourier
transform to evaluate this in the limit $x \to 0^+$, we obtain
\begin{equation}
u(0) = \frac{\Delta}{\sqrt{1+k}} - \frac{k\eta v}{1+\sqrt{1+k}} u(0)
\end{equation}
Recognizing $u(0)=\epsilon$ and $\sqrt{1+k}= \Delta_G/\epsilon$, after
reorganizing we obtain
\begin{equation}
\eta v=\left(\frac{\Delta}{\Delta_G} -1\right)\frac{1 + \sqrt{1+k}}{k}
\end{equation}
which is easily seen to be equivalent to what we obtained from the
direct method.  As can be appreciated, the Wiener-Hopf method is much
more involved than the direct method.  Nevertheless, it will be the essential
tool for analyzing the large-$N$ limit.

\section{General $N$}

It is straightforward to extend the lattice calculations to arbitrary
N.  The basic method is the same: we solve the problem on the
two sides of the crack tip position and patch the two solutions together.
The solution on either side is again a sum over modes, which are a direct
product of modes in the vertical direction, given by the eigenmodes of
${\cal M}_N(m=0,k)$, with the modes in the horizontal direction.  Thus there are
a total of $Nn_u$ and $Nn_c$ modes on the uncracked and cracked sides,
respectively. The solutions on either side have to overlap for each
value of the vertical component $j$, so there are an appropriate number
of equations for the unknown coefficients of each mode.  As for $N=1$,
the condition $u(0,1)=\epsilon$ is used to determine the driving $\Delta$
corresponding to a given velocity.

\begin{figure}
\centerline{\epsfxsize=3.25in \epsffile{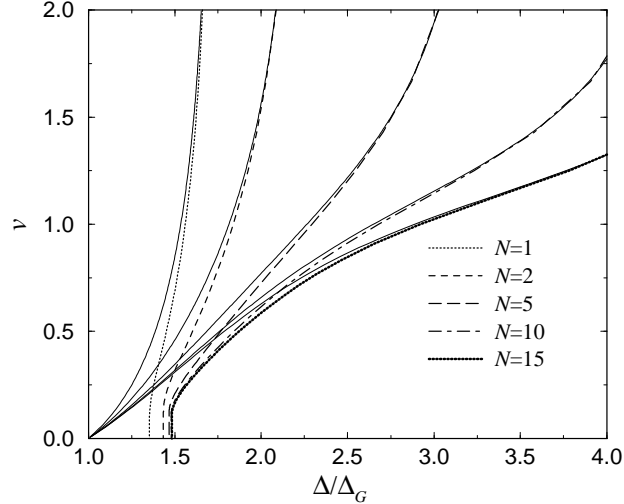}}
\caption{$v$ versus $\Delta/\Delta_G$ for $\eta=2$ for $N=1,\ 2,\ 5,\ 10,\ 15$;
$k=2$. The calculation was done with $dt=.1$.  
The solid lines are the naive continuum results for the same
parameters.}
\label{fig4}
\end{figure}

We can also generalize our continuum calculation to finite $N$.  As in 
the $N=1$ case, we replace finite differences in $t$ with derivatives, giving
us $N$ coupled third-order differential equations.  Again, we can solve
these exactly on either side of the crack tip $t=0$, and match the functions
and their first and second derivatives at this point.  The functions
are characterized by $N$ modes on the uncracked side, with decay rates
$Q_{1,n}$, and $2N$ modes on the cracked side, with decay rates $q_{2,n}$,
$q_{3,n}$.  For each $n$, $Q_{1,n}$ is the positive root of 
of the polynomial $P({\Lambda_n})$ defined in Eq. (\ref{Qpoly}) above.
Let us denote the other roots of this polynomial, which we will
need later, by $-Q_{2,n}$, $-Q_{3,n}$.  Similarly, $-q_{2,n}$, $-q_{3,n}$
are the two negative (real part) roots of $P(\lambda_n)$. 
with the third, positive, root being labeled by $q_{1,n}$.
Implementing these procedures, we again calculate the crack
velocity as a function of $\Delta/\Delta_G$.  Again, we compare this
data to that of our naive continuum (in $x$) calculation for the same
value of $N$.  We present in Fig. \ref{fig4} the results for our
overdamped case $\eta=2$, for $N=1,2,5,10,15$.  Qualitatively, not much
changes with $N$.  The most important feature in that the middle section
of the curves get progressively flatter as $N$ increases.  This must
be the case, since the point of divergence, $\Delta_U$, measured in terms
of $\Delta_G$, increases as $N^{1/2}$.  We also note that the data for
low velocities seems to converge fairly rapidly as $N$ increases, and
the rate of convergence slows as $v$ increases.  Again, as in the $N=1$
case, the continuum results accurately reproduce the lattice calculations
for large $v$ and are completely wrong for small $v$, missing the
lattice-induced arrest phenomenon. 

\section{Large N}
\label{sec:large-n}
The physical problem of cracking a macroscopic object corresponds to
the limit of large, but finite, $N$.  The lattice calculations are
prohibitively expensive for too large $N$.  However, our naive continuum
calculations can be carried out for fairly large $N$'s.  Using the
fact that for small $v$, the convergence in $N$ is rapid, and for
larger $v$, the naive continuum results are reliable, we can piece together
a fairly complete picture of what we expect in the macroscopic limit.
In particular, it is interesting to compare this with the standard
continuum theory in order to understand the limitations and successes
of that theory.

\begin{figure}
\centerline{\epsfxsize=3.25in \epsffile{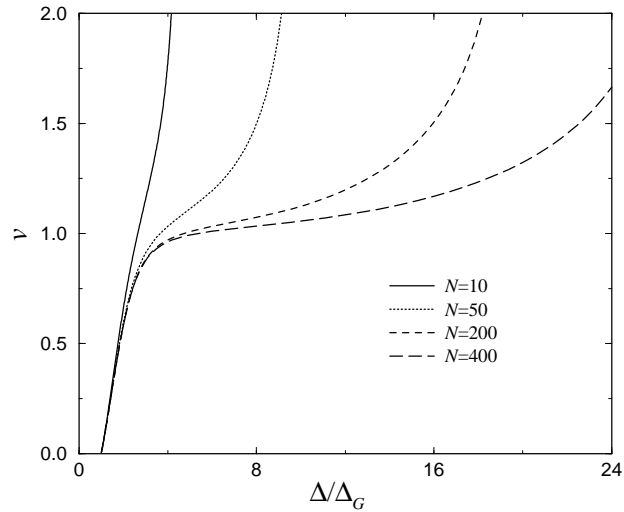}}
\caption{$v$ versus $\Delta/\Delta_G$ in the continuum approximation
for for $N=10,\ 50,\ 200,\ 400$, with $\eta=2$, 
$k=2$. }
\label{fig5}
\end{figure}

To begin, we present in Fig. \ref{fig5} the results of our naive 
continuum theory, extended to larger values of $N$.  The most striking
feature of this graph is the slow convergence that sets in near $v=1$.
Exploring numerically, we find that for fixed $v<1$, the data converges
with large $N$ as $N^{-1/2}$.  However, the coefficient of this $N^{-1/2}$
correction becomes ever larger as $v$ approaches unity.  Looking at
the value of $\Delta/\Delta_G$ where $v=1$, it appears to be diverging
as $N^{1/6}$ as $N \to \infty$.  Thus, in the macroscopic limit,
the crack speed is effectively bounded by the wave speed.

To proceed further in studying our naive continuum theory at large
$N$, it is useful to derive the Weiner-Hopf solution.  To do this,
we first Fourier transform the fields, writing
\begin{equation}
u_j(t)=\int_{-\infty}^{\infty} \frac{dk}{2\pi} 
e^{-ivKt} {\tilde u}_j(K)
\end{equation}
The equations for all $j\ne 1$
are translationally invariant in $t$, and become algebraic. 
The structure of these equations is
\begin{equation}
0=\left(i\eta v K^3 - (1 - v^2)K^2\right){\tilde u}_j 
+ (1 - i\eta Kv){\cal M}_{N-1;j,j'}(1)
{\tilde u}_{j'} + \delta_{j,2} (1-i\eta vK){\tilde u}_1 +
\delta_{j,N}\Delta\delta(K)
\end{equation}
Defining $f(K)\equiv(i\eta v K^3 - (1-v^2)K^2)/(1-i\eta v K)$ and denoting
the $n\times n$ identity matrix as ${\cal I}_n$, we can using Cramer's rule
explicitly solve for ${\tilde u}_2$ in terms of ${\tilde u}_1$ as follows
\begin{equation}
{\tilde u}_2 = -\frac{det_{N-2} \left(f(K){\cal I}+{\cal M}(1)\right)}{det_{N-1} \left(f(K){\cal I}+{\cal M}(1)\right)}
{\tilde u}_1 - \frac{(-1)^N\Delta}{det_{N-1} {\cal M}(1)}\delta(K)  \; .
\end{equation}
where in the last term we have used the $\delta (K)$ to simplify the 
determinant.  
To treat the $j=1$ with its step functions we define
\begin{equation}
{\tilde u}^\pm = \int_{-\infty}^{\infty}\,vdt \ \theta(\pm t)
e^{iKvt} u_1(t)
\end{equation}
so that ${\tilde u}_1 = {\tilde u}^+ + {\tilde u}^-$. 
The $j=1$ equation now reads
\begin{equation}
0  =  (1-i\eta v K) \left[ (f(K)-1){\tilde u}_1  -  
\frac{det_{N-2} \left( f(K){\cal I}+{\cal M}(1) \right) }
{det_{N-1} \left( f(K){\cal I}	+ {\cal M}(1) \right) } {\tilde u}_1 
- k{\tilde u}^- \right] \nonumber \end{equation} \begin{equation} 
\;\;\;\;\;\;\; - k\eta v  u_1(0) - 
\frac{(-1)^N\Delta} {det_{N-1} {\cal M}(1)} \delta(K)
\end{equation}
Multiplying through by $det_{N-1} \left({f(K){\cal I} +\cal M}(1)\right)/
(1-i\eta vK)$, we get 
\begin{eqnarray}
\label{ueq1}
0 &=& det_N(f(K){\cal I} + {\cal M}(0)){\tilde u}^-  + det_N(f(K){\cal I} + {\cal M}(k)){\tilde u}^+ \nonumber\\
&\ &\ - \frac{k\eta v u_1(0)det_{N-1}(f(K){\cal I}+{\cal M}(1))}{1-i\eta v K}
 - (-1)^N\Delta\delta(K)
\end{eqnarray}
The determinants are easy to calculate in the diagonal bases of the $\cal M$'s,
and have zeros at $K$'s corresponding precisely
to $i$ times the roots of the polynomials $P(\Lambda_n)$, $P(\lambda_n)$
we encountered in our original real-space calculation.  We can
thus write
\begin{eqnarray}
det_N(f(K){\cal I} + {\cal M}(k)) &=&\left(1-i\eta v K)\right)^{-N}
\prod_{n=1}^N P(\Lambda_n;-iK) \nonumber\\
&=&\left(\frac{i\eta v}{1-i\eta v K}\right)^N
\prod_{n=1}^N (K-iQ_{1,n})(K+iQ_{2,n})(K+iQ_{3,N})
\end{eqnarray}
and similarly
\begin{eqnarray}
det_N(f(K){\cal I} + {\cal M}(0)) &=&
(1-i\eta v K)^{-N}\prod_{n=1}^NP(\lambda_n;-iK) \nonumber\\
&=&\left(\frac{i\eta v}{1-i\eta v K}\right)^N 
\prod_{n=1}^N (K-iq_{1,n})(K+iq_{2,n})(K+iq_{3,N})
\end{eqnarray}
Similarly, if we denote the eigenvalues of ${\cal M}_{N-1}(1)$ by $\ell_m$, 
$m=1,\ldots,N-1$, we can express $det_{N-1}(f(K){\cal I}+{\cal M}(1))$ in terms of
the roots $\chi_{1,m}$, $-\chi_{2,m}$, and $-\chi_{3,m}$ of $P(\ell_m)$
\begin{eqnarray}
det_{N-1}(f(K){\cal I} + {\cal M}(1)) 
&=&(1-i\eta v K)^{-N-1}\prod_{m=1}^{N-1}P(\ell_m;-iK) \nonumber \\
&=&\left(\frac{i\eta v}{1-i\eta v K}\right)^{N-1}
\prod_{m=1}^{N-1} (K-i\chi_{1,m})(K+i\chi_{2,m})(K+i\chi_{3,m}) \;.
\end{eqnarray}
We can then re-express Eq. \ref{ueq1} as
\begin{eqnarray}
0 &=& \prod_n\frac{K-iQ_{1,n}}{K-iq_{1,n}}{\tilde u}^-
 + \prod_n\frac{(K+iq_{2,n})(K+iq_{3,n})}{(K+iQ_{2,n})(K+iQ_{3,n})}{\tilde u}^+ \\
&\;&\; + \frac{ik\prod_m(K-i\chi_{1,m})(K+i\chi_{2,m})(K+i\chi_{3,m})}
{\prod_n(K-iq_{1,n})(K+iQ_{2,n})(K+iQ_{3,n})}u_1(0)
- \frac{\Delta}{(\eta v)^N\prod_n q_{1,n}Q_{2,n}Q_{3,n}}\delta(K) \nonumber\\
 &=& \prod_n\frac{K-iQ_{1,n}}{K-iq_{1,n}}{\tilde u}^-
 + \prod_n\frac{(K+iq_{2,n})(K+iq_{3,n})}{(K+iQ_{2,n})(K+iQ_{3,n})}{\tilde u}^+ \\
&\;&\; + \frac{ik\prod_m(K-i\chi_{1,m})(K+i\chi_{2,m})(K+i\chi_{3,m})}
{\prod_n(K-iq_{1,n})(K+iQ_{2,n})(K+iQ_{3,n})}u_1(0)
- \Delta\prod_n\frac{q_{2,n}q_{3,n}}{Q_{2,n}Q_{3,n}}\delta(K)
\end{eqnarray}
where in the last step, we applied the identity
\begin{equation}
\label{prodq}
\prod_n q_{1,n}q_{2,n}q_{3,n}=
(-\eta v)^{-N}det {\cal M}_N(0) = (\eta v)^{-N} \ . 
\end{equation}
and where we have used the easily verified fact that 
$det {\cal M}_N (m)=(-1)^{N}(mN+1)$.
Note that this product result nicely reduces to the result we previously 
obtained for $N=1$.
Again, as in the $N=1$ case, to proceed with the Wiener-Hopf method, 
we need to break up the last
two terms into pieces analytic in the upper and lower half-planes.  The
$u_1(0)$ piece does not appear to have a simple breakup.  However, 
for large $N$, the
effect of this term becomes irrelevant, since $u_1(0)$ is a factor $N^{1/2}$
smaller than $\Delta$.  The last term is easily broken up as we did in the
$N=1$ case. We find that to leading order
\begin{equation}\label{finalformula}
{\tilde u}^+ \approx \Delta \left(\frac{i}{K+i0^+}\right)\prod_n
\frac{q_{2,n}q_{3,n}(K+iQ_{2,n})(K+iQ_{3,n})}
{Q_{2,n}Q_{3,n}(K+iq_{2,n})(K+iq_{3,n})} \ .
\end{equation}

In the large $N$ limit, we can use this to evaluate $u$ explicitly.
(We need not concern ourselves with ${\tilde u}^-$, since
for $t<0$, $u_1(t)$ is always smaller that $u_1(0)$, and so does not
contribute to leading order.) To proceed, we need the explicit form
of the $q,Q$'s to leading order.  As we shall see, the behavior is
controlled by modes where $n<<N$.  For these modes, we may approximate
the eigenvalues of ${\cal M}_N (k)$ by $\Lambda_{1,n}=\Lambda_{2,n}=-n^2\pi^2/N^2$, so that $Q_{1,n}=Q_{2,n}=
\frac{n\pi}{N\sqrt{1-v^2}}$, $Q_{3,n}=(1-v^2)/(\eta v)$. (Here we
have assumed that $v$ is less than and not too close to $1$.) Similarly, for
${\cal M}_N (0)$, we find
$\lambda_{1,n}=\lambda_{2,n}=-(n-\frac{1}{2})^2\pi^2/N^2$, so that 
$q_{1,n}=q_{2,n}= \frac{(n-\frac{1}{2})\pi}{N\sqrt{1-v^2}}$, 
$q_{3,n}=(1-v^2)/(\eta v)$. Notice that since $Q_{3,n} \approx q_{3,n}$,
the factors involving these quantities cancel.
This has the immediate consequence that the viscosity $\eta$ has
completely dropped out of the problem in this limit.

The
remaining expression has poles at $-i0^+$ and at $-iq_{2,n}$.  We
can evaluate the residue of each of these poles explicitly. The residue at
$-i0^+$ is immediately seen to be
\begin{equation}
Res (e^{-iKvt}{\tilde u}^+)|_{-i0^+} = i\Delta \ .
\end{equation}
Evaluating the residues at the other poles
is more complicated. To proceed,
let us cut-off the product at some large $n \equiv N_c << N$. Then, our approximate
expressions for the $q$'s and $Q$'s are valid. This leads to
\begin{eqnarray}
Res (e^{-iKvt}{\tilde u}^+)|_{-iq_{2,n}} &=&
-i e^{-\frac{(n-\frac{1}{2})\pi v t}{N\sqrt{1-v^2}}} 
\frac{\Gamma (N_c+\frac{1}{2}) \Gamma (n-\frac{1}{2}) 
\Gamma (N_c+\frac{3}{2}-n)}
{(n-\frac{1}{2})\Gamma (N_c +1) \Gamma^3(\frac{1}{2}) \Gamma (n) 
\Gamma (N_c+1-n)} \nonumber \\
&\simeq& -ie^{-\frac{(n-\frac{1}{2})\pi v t}{N\sqrt{1-v^2}}} 
\frac{\Gamma (n-\frac{1}{2})}{(n-\frac{1}{2}) \pi^{\frac{3}{2}}
\Gamma (n)} \ + \ O(\frac{1}{N_c})
\end{eqnarray}
We then take the limit of $N_c \rightarrow \infty$ to find the final answer
for the macroscopic displacement field
\begin{eqnarray}
\label{arcsin}
u_1(t)&=&\theta(t) \Delta \left[ 1 \ - \ \sum_{n=1}^{\infty}
\frac{\Gamma (n-\frac{1}{2})}{(n-\frac{1}{2}) \pi^{\frac{3}{2}}
\Gamma (n)} e^{-\frac{(n-\frac{1}{2})\pi v t}{N\sqrt{1-v^2}}} \right] \nonumber \\
& = &\theta(t) \Delta \left( 1 - \frac{2}{\pi} \sin^{-1}
{e^{-\frac{\frac{1}{2}\pi v t}{N\sqrt{1-v^2}}}} \right)
\end{eqnarray}

\begin{figure}
\centerline{\epsfxsize=3.25in \epsffile{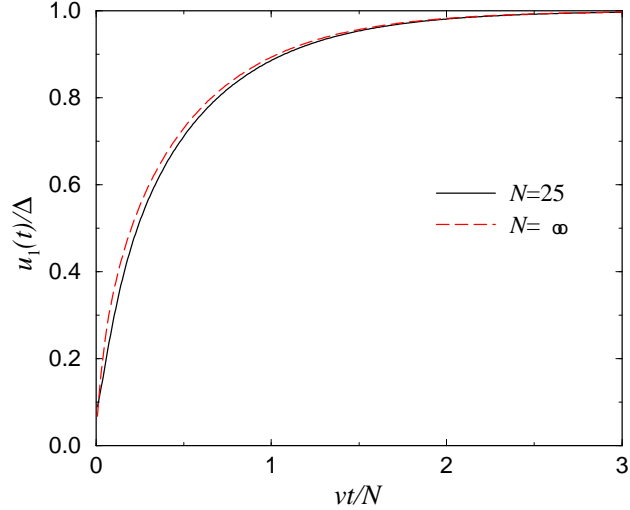}}
\caption{$u_1(t)/\Delta$ versus $vt/N$ in the continuum approximation
for $k=2$, $\eta=2$, $N=25$, compared with the large-$N$ analytic
result, Eq. (\protect\ref{arcsin}).
}
\label{fig6}
\end{figure}

This final answer exhibits the well-known square root branch cut at
the crack tip location, $t=0$. It is worth noting that this behavior 
of the displacement gives rise to
a macroscopic stress field which actually diverges as $\frac{1}{x^{3/2}}$
(recall the extra derivative due to the Kelvin viscosity) near the
crack tip. This surprising finding renders invalid the 2-d continuum
calculation of Langer\cite{langer} who studied this problem with the additional
complication of a finite length cohesive zone. A correct continuum formulation
which does reproduce the essential formula Eq. (\ref{finalformula}) is
presented in the Appendix. 

\begin{figure}
\centerline{\epsfxsize=3.25in \epsffile{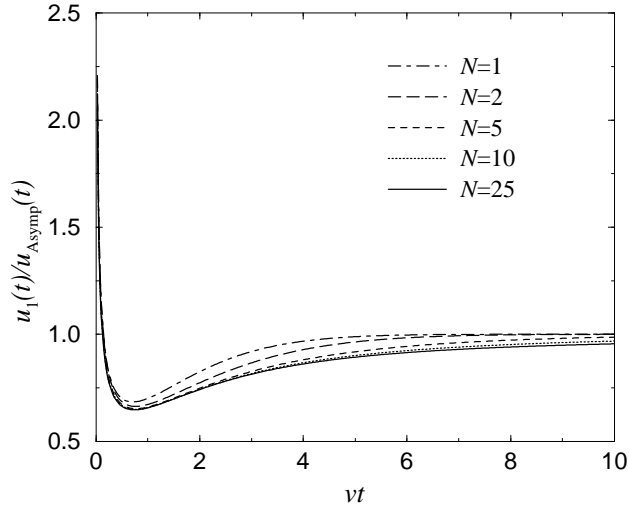}}
\caption{$u_1(t)/u_{\text Asympt}(t)$ versus $vt$ in the 
continuum approximation where $u_{\text Asympt}$ is the large-$N$
analytic result, Eq. (\protect\ref{arcsin}), 
for $N=1, 2, 5, 10, 25$.  Again, $k=2$, $\eta=2$.
}
\label{fig7}
\end{figure}

A comparison of the above prediction
with the numerically computed displacement is shown in Fig. (\ref{fig6})
for the case $N=25$, plotted as a function of the macroscopic scale $vt/N$.  
We see that our large $N$ analytic result correctly
reproduces the large-scale structure of the crack displacement.  It
does worse, though still quite accurate, close in to the crack tip.  To
demonstrate this more quantitatively, we present in Fig. (\ref{fig7})
the ratio of the crack displacement $u_1(t)$ to the large-$N$ analytic result,
for various $N$. Now the data is plotted as a function of the
microscopic scale $vt$.
We see that curves are all quite similar.  They
have a square-root divergence at the origin, since the analytic prediction
is that $u(t)$ vanishes at $t=0$, whereas the true answer is finite.
By $N=25$, they have converged to a limiting curve.  This means that
the finite-$N$ theory possesses a well-defined microscopic structure,
in addition to the universal macroscopic structure defined by the
standard continuum theory.  This microscopic structure is on the
scale of the lattice constant (in the $y$-direction) and is of course
invisible to the standard continuum theory.

This observation implies that
the entire issue of velocity selection via the condition that $u(x=0,y=1)$
be fixed to equal the breaking displacement is out of reach of the leading
order macroscopic limit. Thus, as an example, the velocity depends 
explicitly on $\eta$
(as opposed to the macroscopic displacement which explicitly does not) even
for arbitrarily large $N$. Conversely, calculating the macroscopic
field in the continuum limit does not suffice for determining the crack speed
which is always fixed at the lattice scale. Situations with  equivalent
macroscopic fields can have arbitrarily different crack velocities.

\section{Summary}

In this paper, we have studied in some detail the steady-state motion
of mode III viscoelastic cracks in a lattice model of the microscopic
dynamics. The most important finding are
\begin{enumerate} 
\item The existence of a minimum velocity for crack propagation is
dependent on the viscosity. At low $\eta$ (and indeed in the
lattice models without dissipation that have been studied to date),
the steady-state branch starts at finite $v$. For highly damped systems,
on the other hand, the branch extends all the way to $v=0$ at the
upper end of the allowed range of $\Delta$ for arrested cracks.
\item For finite $N$, a continuum approach (in $x$) for the crack
does accurately predict the lattice results for values of the driving away from
the lattice trapping (low or zero velocity) regime. 
\item Taking the macroscopic limit ($N \rightarrow \infty$) allows us
to recover the expected macroscopic behavior that the displacement
grows as $\sqrt{x-x_{\mbox {tip}}}$ once we leave an inner core region
of the lattice scale. The coefficient of this term can be calculated
by using a continuum theory with the proper boundary conditions. A key feature
of this macroscopic theory is that the viscosity becomes irrelevant. 
\item However, the velocity selection as a function of
the imposed displacement is wholly controlled by the core and cannot be
accurately arrived at by any theory which does not explicitly consider
the lattice scale. In particular, viscosity plays a crucial role in
this feature of the physics.
\end{enumerate}

As mentioned in the Introduction, the next step in our research program
will be to consider the modifications introduced into the aforementioned
results by having a continuous but nonlinear force law. In particular,
having a force which immediately drops to zero means that there is no
way that the system could dynamically decide to create a ``cohesive"
zone of mesoscopic (i.e., scaling as a positive power of $N$) proportions.
In such a zone the displacement would be such that the force law would
be beyond the linear spring regime but not so large that the force would
be effectively zero). If it were of large size, it would lead to a
more complex continuum theory along the lines suggested by Langer and
co-workers~\cite{langer,langer+others}; if it were purely on the lattice
scale, it would change nothing. Our initial evidence suggests the latter,
and we hope to report on this in the near future.

As far as the physics of fracture is concerned, we must address several
issues that go well beyond the studies in this paper. Since
most of the experiments concern mode I cracks, we need to extend our
results to that situation; this is technically challenging but should
not lead to any significant surprises. Next, we must explicitly investigate
the stability of our steady-state equations. Finally, all lattice models
leave out the possibility of ductile behavior involving the emission of
dislocations from the crack tip; comparison to experiment and to
molecular dynamics simulations will enable us to learn when these additional
phenomena are crucial or alternatively  when one can get by with a
purely ``brittle" model.

\acknowledgments
HL acknowledges the support of the US NSF under grant DMR98-5735; DAK 
acknowledges the support of the Israel Science Foundation and the
hospitality of the Lawrence Berkeley National Laboratory. The work
of DAK was also supported in part by the Office of Energy Research,
Ofice of Computational  and Technology Research, Mathematical, Information 
and Computational Sciences Division, Applied Mathematical Sciences Subprogram,
of the U.S. Department of Energy, under Contract No. DE-AC03-76SF00098.
Also, DAK acknowledges useful conversations with L. Sander and M. Marder.

\appendix
\section*{The Direct Continuum Calculation}
In this appendix, we present a direct continuum (in $x$ and $y$)
calculation of the steady-state crack.  We will see that it recovers 
directly the leading-order results of the large-$N$ limit calculation 
presented in section \ref{sec:large-n} above. 

To begin, we write the displacement field $u(x,y,t)=u(t-x/v,y)$ 
in Fourier space:
\begin{equation}
u(t,y)=\Delta\frac{y}{W} + \int_{-\infty}^\infty \frac{dK}{2\pi} {\tilde u}(K)
e^{-iKvt}\frac{\sinh(k_y (W-y)}{\sinh(k_y W)}
\end{equation}
where $k_y$ satisfies the dispersion relation
\begin{equation}
(1-i\eta v K) (-K^2 + k_y^2) + v^2 K^2 = 0
\end{equation}

The crack is chosen to begin at $x=0$ so $u(x<0,0)=0$. On the crack 
surface $y=0$, $x>0$, we must set $du/dy=0$. Note that this condition
implies that the normal stress on the free surface, $(1+\eta v \frac{d}{dx})
\frac{du}{dy}$ vanishes.  However, it is incorrect to assume, 
as Langer~\cite{langer} did in a parallel calculation, that the
vanishing of the normal stress is a {\em sufficient} condition, as this allows
for (unphysical) displacement fields that do not have $du/dy=0$.
As we have seen, the macroscopic field possesses a square-root singularity
at the origin, while Langer's condition eventually results in a much weaker $x^{3/2}$
singularity (in the absence of Barenblatt type surface stresses). Our
condition implies
\begin{equation}
\int_{-\infty}^\infty \frac{dK}{2\pi} {\tilde u}(K) (-k_y)\coth(k_y W) =
-\frac{\Delta}{W} + \theta(-t)G(t)
\end{equation}
or, Fourier-transforming this equation:
\begin{equation}
\label{cont-wh}
{\tilde u}(K) (-k_y W)\coth(k_y W) = \frac{-\Delta}{W}\delta(K)
+ {\tilde G}^-(K)
\end{equation}
where ${\tilde G}^-$ is the transform of $\theta(-t)G$ and has no
zeros or roots in the lower-half-plane. To proceed, we use the identity
\begin{equation}
k_y W \coth(k_y W) = \prod_{n=1}^{\infty} \frac{
1+\left(\frac{k_yW}{(n-\frac{1}{2})\pi}\right)^2}
{1+\left(\frac{k_yW}{n\pi}\right)^2} 
\end{equation}
Now, we can use the dispersion relation to eliminate $k_y$ in favor of
$K$. If we define $\lambda_n \equiv -[(n-\frac{1}{2})\pi/W]^2$ and
$\Lambda_n \equiv -(n\pi/W)^2$, then we find that
\begin{equation}
k_y W \coth(k_y W) = \prod_{n=1}^{\infty} \frac{
\Lambda_n P(\lambda_n;-iK)}{\lambda_n P(\Lambda_n;iK)}
\end{equation}
Notice that $\Lambda_n$, $\lambda_n$ are precisely the same as those
in the finite-$N$ calculation for $n \ll N$, if we identify $W=N$.  
Expressing the $P$'s in terms of their roots, we get
\begin{equation}
k_y W \coth(k_y W) = \prod_{n=1}^{\infty} \frac{
Q_{1,n}Q_{2,n}Q_{3,n} (K-iq_{1,n})(K+iq_{2,n})(K+iq_{3,n})}
{q_{1,n}q_{2,n}q_{3,n} (K-iQ_{1,n})(K+iQ_{2,n})(K+iQ_{3,n})}
\end{equation}
Plugging this into Eq. (\ref{cont-wh}), and reorganizing, we obtain
\begin{equation}
{\tilde u}^+ \prod_n\frac{Q_{2,n}Q_{3,n}(K+iq_{2,n})(K+iq_{3,n})}
{q_{2,n}q_{3,n}(K+iQ_{2,n})(K+iQ_{3,n})} + 
{\tilde u}^- \prod_n\frac{q_{1,n}(K-iQ_{1,n})}{Q_{1,n}(K-iq_{1,n})}
= \Delta \delta(K) - W{\tilde G}^- \prod_n  \frac{q_{1,n}(K-iQ_{1,n})}{Q_{1,n}(K-iq_{1,n})}
\end{equation}
Decomposing the $\delta$-function as in the finite-$N$ case, and separating
out the pieces analytic in the upper-half plane, we get
\begin{equation}
{\tilde u}^+ \prod_n\frac{Q_{2,n}Q_{3,n}(K+iq_{2,n})(K+iq_{3,n})}
{q_{2,n}q_{3,n}(K+iQ_{2,n})(K+iQ_{3,n})} 
= \frac{\Delta}{W}\frac{i}{K+i0^+}
\end{equation}
so that
\begin{equation}
{\tilde u}^+ = \frac{i\Delta}{K+i0^+}\prod_n\frac{q_{2,n}q_{3,n}(K+iQ_{2,n})
(K+iQ_{3,n})}{Q_{2,n}Q_{3,n}(K+iq_{2,n})(K+iq_{3,n})} 
\end{equation}
That this result is the direct equivalent of our leading-order 
finite-$N$ result, Eq. (\ref{finalformula}), is clear.  One word
of interpretation is called for, however.  To achieve the macroscopic
limit of our finite-$N$ result, we needed to take the width $N$ large.
This in turn implied that viscosity was irrelevant in the macroscopic
limit (unless we scaled it by a power of $N$ with no obvious physics
justification). If we work directly in the continuum, however, we obtain the
same final result without having to take $W$ large.  Thus, the ratio
of the viscous length scale to $W$ is arbitrary in this continuum calculation.
Nevertheless, if we examine the large-$W$ limit of our continuum calculation,
we again will find that viscosity becomes irrelevant.  It is also worth
reiterating that this continuum calculation has no sign of the subdominant
pieces which control velocity selection.

\end{document}